\documentclass[12pt]{article}
\usepackage{graphicx}
\usepackage[utf8]{inputenc}
\usepackage{amsmath} 
\usepackage{amssymb}

\begin{document}
\begin{center} \Large {\bf ``Palatini's Cousin: A New Variational Principle''}\normalsize\\\large
  {\bf Hubert Goenner} 
  \\ University of G\"ottingen - Institute for Theoretical Physics\\
  Friedrich-Hund-Platz 1 - D 37077 G\"ottingen\normalsize \end{center}
\vspace{1cm}
{\bf Abstract}:\\ A variational principle is suggested within Riemannian
geometry, in which an auxiliary metric and the Levi Civita connection
are varied independently. The auxiliary metric plays the role of a
Lagrange multiplier and introduces non-minimal coupling of matter to
the curvature scalar. The field equations are 2nd order PDEs and easier
to handle than those following from the so-called Palatini
method. Moreover, in contrast to the latter method, no gradients of
the matter variables appear. In cosmological modeling, the physics
resulting from the new variational principle will differ from the
modeling using the Palatini method.\\

PACS numbers: 04.20 Fy, 04.50 Kd, 98.80 Cq, 02.40 Ky

\section{Introduction}
For the derivation of the field equation of Einstein's theory of gravitation 
and of alternative gravitational theories sometimes a method named, 
alternatively, ``Palatini's Principle'', ``the Palatini method of
variation'' or ``Palatini's device'' is used. Although the starting
point is Riemannian geometry, besides the metric an independent
affine connection forming the curvature tensor is imagined; in the
Lagrangian, both metric and connection then are varied independently. An 
advantage of the method is that it leads to 2nd order field equation
for Lagranians of higher order in curvature while a variation of the
metric as the only variable results in 4th-order PDEs. On the other
hand, a main conceptual difficulty of the method is that the variational
procedure mixes Riemannian and metric-affine geometry. Authors either
leave undetermined the space-time geometry as a frame for the new
connection, or tacitely fix it  mentally by introducing constraints (symmetric
connection, no torsion etc) which do not show up in the formalism. 

Since many years, warnings have been voiced that the method be
working reliably only for the Hilbert-Einstein Lagrangian (plus the
matter part) ${\cal L} =  \sqrt{-g}~[R(g_{ij}) + 2\kappa L_{mat}(g_{ij}, u^{A})]$ with curvature scalar $R=g^{lm} R_{lm}(g_{ij})$ and matter 
variables $u^A$, but otherwise leads to under- and
un-determinacies \cite{Buchdahl1960}, \cite{Buchdahl1977},
\cite{Stephen1977}.\footnote{For incorrectly relating Palatini's name with
  what is ascibed to him cf. \cite{Hehl1980}, footnote on  p. 40 as well as
  the English translation of Palatini's paper in the same volume on pp. 477-488
  (1980). Cf. also \cite{Ferra1982}.} Recently, Palatini's method has been 
unearthed in attempts to build cosmological models thought to explain the
accelerated expansion of the universe with its consequences for dark
energy \cite{Voll03}, \cite{Flana04}, \cite{VaTaTsu07}, \cite{SotLib07},
\cite{Olmo08}. The method also has been applied to loop quantum cosmology
\cite{OlmSin09}. Often, the starting point is a Lagrangian of the form ${\cal L}
= \sqrt{-g}~[~R(g_{ij}) + \tilde{f}(R)~] +  \sqrt{-g}~ 2\kappa L_{mat}(g_{ij}, 
u^{A})$ with $\tilde{f}$ an arbitrary smooth function.\footnote{Recently, 
Lagrangians with two curvature invariants, i.e., $f(R,~
R_{ab}R_{lm}g^{al}g^{bm})$ have been considered \cite{OlmSanTri09}.} In the 
following, we suggest another variational principle leading to 2nd order field 
equations and lacking the deficiencies of the Palatini method. After its
introduction, it is applied to the class of $f(R)$-theories in section
\ref{section:ftheory} and compared with the Palatini method in section 
\ref{section:compalat}. A recent particular choice for $f(R)$ in the framework 
of cosmological modeling then is used as an example for the working of the new 
principle.

\section{The new variational principle}
\label{section:newvar}
Whereas in the Palatini method the Levi Civita connection
(represented by the Christoffel symbol) is replaced by a general
affine connection, here we keep the geometry (pseudo-)Riemannian but
introduce an auxiliary Lorentz metric. This is done by replacing, in an action 
integral set up within Riemannian geometry, the (Lorentz-)metric $g_{ab}$ by
an auxiliary metric $\gamma_{ab}$ except in the Levi Civita connection which
is left unchanged. The independent variables for the variation are
$\gamma_{ab}$ and the Levi Civita connection formed from $g_{ab}$ 
\begin{equation}\{ _{ij}^{k}\}_g = \frac{1}{2}g^{kl}(\frac{\partial
    g_{il}}{\partial x^j}+\frac{\partial g_{jl}}{\partial x^i} -
  \frac{\partial g_{ij}}{\partial x^l})~\label{christo}.\end{equation}
The equations following from the variation will give
the dynamics of the gravitational field and link $\gamma_{ab}$ with $
g_{ab}$. We wish to emphasize that it is {\em not} a  bi-metric theory which is 
aimed at.\footnote{In bi-metric theories, one metric usually
  is fixed to be the flat Minkowskian metric and not varied. A formal
  variation of the second metric often is restricted to an infinitesimal 
coordinate change in order to derive conservation
laws. Cf. \cite{Kohler1952}.} The auxiliary metric may be seen as playing the
role of a Lagrange multiplier. This is analogous to the case of scalar-tensor 
theories replacing $f(R)$-theories of gravitation (cf. \cite{Soti06}, 
\cite{Igle07}). The new variational principle for Einstein gravity starts 
from:\footnote{Latin indices a, b, i, j, ... run from 0 to 3; the summation 
convention is implied. } \begin{equation}{\cal L} =
  \sqrt{-\gamma}~[\gamma^{ab}~ R_{ab}(\{ _{ij}^{k}\}_{g}) + 2\kappa
  L_{mat}(\gamma_{lm}, u^A)]~.\label{lagr1}\end{equation}  Variation with
  respect to $\gamma^{ab}$ leads to:
 \begin{equation}{\delta_{\gamma}\cal L} =\sqrt{-\gamma}~[~R_{ab}(\{
    _{ij}^{k}\}_g) - \frac{1}{2}\gamma_{ab}~R_{\gamma}
    + 2\kappa~T_{ab}(\gamma_{lm}~, u^A)~]~\delta \gamma^{ab} \label{new1}, 
\end{equation} where $R_{\gamma}:= \gamma^{lm}R_{lm}(\{_{ij}^{k}\}_g)$ and 
$T_{ab}:= \frac{2}{\sqrt{-\gamma}}~\frac{\delta{\cal L}_{mat}}{\delta
   \gamma^{ab}}$. Variation with respect to $\{ _{ij}^{k}\}_g$ gives: 
\begin{equation}{\delta_{\{ _{ij}^{k}\}_g }\cal L} =
   [~-(\sqrt{-\gamma}~\gamma^{b(i})_{;b}\delta_k^{~j)}+
 (\sqrt{-\gamma}~\gamma^{ij})_{;k}~]~\delta(\{ _{ij}^{k}\}_g) \label{new2} \end{equation} up to divergence terms.\footnote{($\sqrt{-\gamma}A^k)_{;k}$ always may be
   written as $\sqrt{-g}~(\sqrt{\frac{\gamma}{g}}A^k)_{;k}$ and thus
   as ($\sqrt{-\gamma}A^k)_{,k}$.} From ${\delta_{\{ _{ij}^{k}\}_g
   }\cal L}=0$, after a brief calculation using the trace of (\ref{new2}),
 \begin{equation}
   (\sqrt{-\gamma}~\gamma^{ij})_{;k}=0\label{det met} \end{equation}
 follows, where the covariant derivative is formed with the Levi
 Civita connection. Thus,  $\gamma^{ab} = const \cdot g^{ab}$ follows. 
${\delta_{\gamma}\cal L}=0$ from (\ref{new1}) reduces to Einstein's field
 equations.\\

The method is particularly well suited to a calculus with differential
forms. Here, the usual basic 1-forms $\theta^{i}=
e^{i}_{~r}~ dx^r$ and the curvature 2-form $\Omega_{ij}= \frac{1}{2}
R_{ijkl}(g_{lm})\theta^{k} \wedge \theta^{l}$ are taken as the independent
variables. In place of the auxiliary metric $\gamma_{ij}$, now an auxiliary
1-form is introduced and denoted by $\bar{\theta}^{i}= \bar{e}^{i}_{~r}~ dx^r$
where \begin{equation} \bar{e}^{i}_{~r}~ \bar{e}^{j}_{~s}~\eta^{ij}
  =\gamma^{rs},~ e^{i}_{~r}~e^{j}_{~s}~\eta^{ij} = g^{rs}~. \end{equation}  
The Einstein-Hilbert Lagrangian is $ {\cal L}_E = \Omega_{ab}\wedge
*(\bar{\theta}^{a}\wedge \bar{\theta}^{b})$ with the Hodge-star operation:
$*(\bar{\theta}^{a}\wedge \bar{\theta}^{b})=: \bar{\epsilon}^{ab}$ and
$\bar{\epsilon}_{ab}:=  \frac{1}{2!}\epsilon_{ablm}\bar{\theta}^{l}\wedge 
\bar{\theta}^{m}$.\footnote{Notation here is somewhat ambiguous: e.g., the 
curvature form depends on both the Levi Civita connection and the auxiliary
  tetrad: $\Omega_{ij}= \frac{1}{2} R_{ijkl}(\{ _{ij}^{k}\}_g)~\bar{\theta}^{k}
  \wedge \bar{\theta}^{l}$. Nevertheless, no bar  will be put on
  $\Omega$. The notation $\Omega_{ij}(g,\bar{\theta})$ would be
  inconvenient.} Variation with regard to the fundamental 1-forms and curvature
form leads to the field equations:\begin{eqnarray} D(\frac{\partial{\cal
      L}_E}{\partial \Omega_{ij}}) = 0,~~ \frac{\partial{\cal
      L}_E}{\partial \bar{\theta}_{i}} = 0~ \end{eqnarray} with the covariant
external derivative $D$ using the Levi Civita connection (1-form). Because of 
$\frac{\partial{\cal L}_E}{\partial \Omega_{ij}}= 
*(\bar{\theta}^{i}\wedge \bar{\theta}^{j})$ and of $ \frac{\partial{\cal
      L}_E}{\partial \bar{\theta}_{i}}= \Omega_{lm}\wedge \bar{\epsilon}_{ilm}~,
$ the field equations are:\begin{equation} D\bar{\epsilon}^{ij}=0~,~
  \Omega_{lm}\wedge \bar{\epsilon}_{ilm}=0~,\label{fieldeqforms}
\end{equation} where $ \bar{\epsilon}_{ilm}:= \epsilon_{ilmp}\theta_{p}$ is a 
1-form; $\bar{\epsilon}^{ilm}$ is dual to $ \bar{\theta}^{i}\wedge
\bar{\theta}^{l}\wedge \bar{\theta}^{m}.$ Standard manipulations with the
forms show that the 1st equation (\ref{fieldeqforms}) is satisfied
identically due to the absence of torsion, i.e, $D \bar{\theta}^{m}=0$; and that the 2nd becomes: $2 G^{c}_{~a}(g)\bar{\epsilon}_{c}=0$ with
the Einstein tensor $G^{c}_{~a}(g)$ and the 3-form $\bar{\epsilon}_{i}:=
\frac{1}{3!}\epsilon_{iklm}\bar{\theta}^{k}\wedge \bar{\theta}^{l} \wedge 
\bar{\theta}^{m}~. $ An advantage of this formalism is that it may be adapted
easily to gauge theories.

\section{Extension to f(R)-theories}
\label{section:ftheory}The new variational principle easily applies to the Lagrangian
\begin{equation} {\cal L} =  \sqrt{-\gamma}~[f(\gamma^{lm}R_{lm}(\{
  _{ij}^{k}\}_g)) + 2\kappa~ {\cal L}_{mat} (\gamma_{ij}, u^{A})]~.
\end{equation} The variations lead to:
\begin{equation}{\delta_{\gamma}\cal L} =\sqrt{-\gamma}~[~f'(R_{\gamma})R_{ab}(\{
    _{ij}^{k}\}_g) - \frac{1}{2}\gamma_{ab}~f(R_{\gamma})
    + 2\kappa~T_{ab}(\gamma_{lm}~, u^A, \partial u^A)~]~\delta
    \gamma^{ab} \label{new1f}, \end{equation} whith $f':=
  \frac{df}{dR}$ and to \begin{equation}\delta_{\{ _{ij}^{k}\}_g \cal L} =
   [~-(\sqrt{-\gamma}~f'(R_{\gamma})\gamma^{b(i})_{;b}\delta_k^{~j)}+
 (\sqrt{-\gamma}~f'(R_{\gamma})\gamma^{ij})_{;k}~]~\delta(\{
 _{ij}^{k}\}_g) \label{new2f} \end{equation} up to divergence
terms. As in section \ref{section:newvar}, from (\ref{new2f}) \begin{equation}
   (\sqrt{-\gamma}~f'(R_{\gamma})\gamma^{ij})_{;k}=0\label{detmet}~,\end{equation}
but from which now follows: \begin{equation}
  \gamma^{ab}=f'(R_{\gamma})  g^{ab},~~\gamma_{ab}=(f'(R_{\gamma}))^{-1}
g_{ab}~.\label{metric2} \end{equation} From (\ref{metric2}),
$R_{\gamma}= f'(R_{\gamma}) R_{g}$ with $R_{g}:=
g^{lm}R_{lm}(\{_{ij}^{k}\}_g)$, i.e., the curvature scalar in
(pseudo-)Riemannian space-time. Writing  \begin{equation} R_{g}=
\frac{R_{\gamma}}{f'(R_{\gamma})}=:r(R_{\gamma})~, \label{g-gamma} \end{equation}
the relation $R_{\gamma} = r^{-1}(R_g)$ can be used to remove all
entries of $\gamma_{ab}$ via the curvature scalar in the field
equations following from (\ref{new1f}). Expressed by $g_{ab}$, they read as:
\begin{equation}f'(r^{-1}(R_{g}))~R_{ab}(\{_{ij}^{k}\}_g) -
  \frac{1}{2}~g_{ab}~\frac{f(r^{-1}(R_{g}))}{f'(r^{-1}(R_{g}))} 
    + 2\kappa~T_{ab}((f')^{-1}(r^{-1}(R_{g}))~g_{lm}~, u^A) = 0
    \label{fieldeq2}~.
\end{equation} Equation (\ref{fieldeq2}) shows that, in contrast to 
f(R)-theories  leading to 4th-order differential equations when
derived by variation of only the metric $g_{ab}$, the new field
equations are of 2nd order in the derivatives of $g_{ab}$. The
auxiliary metric is fully determined: $\gamma^{ab}=
f'(r^{-1}(R_g))g^{ab}$; it is not an absolute object. Beyond  acting
as a Lagrange multiplier its main function is
its appearance in the matter tensor causing non-minimal coupling
to the curvature scalar. No further role in the description of the 
gravitational field is played.\footnote{In particular, $\gamma^{ab}$ does {\em not} 
enter the Levi Civita connection, but  only the matter tensor. As a
  metric $\gamma_{ab}$ is incompatible with the Levi Civita connection; its 
non-metricity tensor does not vanish.} For a Lagrangian of the form
  $\sqrt{-g}~[~R(g_{ij}) + \tilde{f}(R)~]$, in the formalism given above $f$
  is to be replaced by $R+\tilde{f}(R), f'$ by $1+\tilde{f}'$ while $f''=
  \tilde{f}'', f'''= \tilde{f}'''$. \\ 

\noindent A.\\ First, a non-vanishing trace (with respect to the auxiliary metric $\gamma$) of
the matter tensor will be assumed $T_{\gamma}:=
\gamma^{lm}T_{lm}(\gamma_{rs}~, u^A)\neq 0$. In this case, the curvature
scalar is seen to be a functional of the trace of the matter tensor. Because of
\begin{eqnarray}T_{\gamma}= f'(R_{\gamma})~ g^{lm} T_{lm}(f'(r^{-1}(R_g))
  g_{rs}, u^A)= f'(r^{-1}(R_g)) T_g(f'(r^{-1}(R_g)) g_{rs}, u^A),
  \label{traceq} \end{eqnarray} with $\tilde{T}_g := g^{lm}
T_{lm}(\gamma_{rs}, u^A)$ from the g-trace of (\ref{fieldeq2}) follows:\begin{equation} f'^2~R_g -2f
  + 2\kappa~f'~\tilde{T}_g= 0~,\end{equation} or, precisely, \begin{equation}
  (f'(r^{-1}(R_g))^2 R_g - 2f(r^{-1}(R_g)) + 2\kappa  f'(r^{-1}(R_g))\tilde{T}_g
  ((f'(R_g))^{-1} g_{lm} u^A) = 0~. \label{traceT}\end{equation} With a
  newly defined function $\omega$ this can be written as \begin{equation} R_g
  = \omega (2\kappa T_g)~,\end{equation} where now $T_g:= g^{lm}T_{lm}
(g_{rs}, u^A)$. From (\ref{traceT}) we conclude that
  (\ref{fieldeq2}) can be cast into the form of Einstein's equations
  with an effective matter tensor. The curvature scalar is coupled directly
  to the matter variables showing up in its trace; no derivatives are
  involved. In fact:\begin{equation}
      R_{ab}(\{_{ij}^{k}\}_g)-\frac{1}{2}~g_{ab}(R_{g})= - \frac{2\kappa}{f'}~[~T_{ab} (\frac{1}{f'}~g_{lm}, u^A) - \frac{1}{2}T g_{ab}~] - \frac{1}{2}g_{ab}~ \frac{f}{(f')^2}~. \label{newEineq} \end{equation}

In the case of {\em perfect fluid} matter with energy density $\mu$ and
  pressure $p$ \begin{equation}
  T_{ab}(\gamma_{rs}, u^A) = (\mu + p)~\gamma_{al}\gamma_{bm}\bar{u}^l
  \bar{u}^m - p~ \gamma_{ab} \label{perfect} \end{equation} with
$\bar{u}^l := \frac{dx^l}{d\bar{s}}$ and $~d\bar{s}^2=\gamma_{lm}dx^l dx^m. $
Hence, $\bar{u}^l= (f')^{1/2}~ u^l~, u^l= \frac{dx^l}{ds} $ and
  \begin{equation} T_{ab}(\gamma_{rs},
  u^A) = (f')^{-1}(R_g)~ T_{lm}(g_{rs}, u^A)~.\end{equation} In this
case, from (\ref{traceq}) a simple relationship for the $\gamma$- and
$g$-traces of the matter tensor follows: \begin{equation}
  T_{\gamma}(\gamma_{rs}~, u^A)= T_g(g_{rs}, u^A)= \mu-3 p~. \end{equation} In
place of $T^{ab}_{~~;~b}=0$ for the Einstein-Hilbert Lagrangian, in this theory
a more general relationship with  $T^{ab}_{~~;~b}\neq 0$ follows from general
covariance. This is also seen by forming the divergence of the Einstein tensor
in (\ref{newEineq}).\\  

\noindent B.\\
For vanishing trace of the matter tensor $T_{\gamma}=0$,
(\ref{traceT}) reduces to \begin{equation} f'(R_{\gamma})R_{\gamma} - 2
  f(R_{\gamma}) = 0~.\label{tracezero} \end{equation} This implies two cases:\\ i)
$f=(f_0 R_{\gamma})^2$, and ii) $f \neq (f_0 R_{\gamma})^2$. The exceptional case 
i) is characterized by an additional scale invariance implying zero
trace for the matter tensor. The field equations (\ref{fieldeq2})
become  \begin{eqnarray} 2(f_0)^2
  (R_{\gamma})[R_{ab}(g)-\frac{1}{4}R_g~g_{ab}] + 2\kappa
  T_{ab}(\frac{1}{2f_0^2R_{\gamma}}~ g_{lm}, u^A) =
  0~,\nonumber \\ R_g = \frac{1}{2(f_0)^2}~,~ T_{\gamma}= T_g
  =0~. \end{eqnarray} If we take a sourceless Maxwell field as matter,
then\begin{equation} T_{ab}(\gamma_{lm}, F_{lm})= \gamma^{lm}F_{al}F_{bm} -
  \frac{1}{4}\gamma_{ab}\gamma^{il}\gamma^{jm} F_{il}F{jm} = f'(R_{\gamma})
  T_{ab}(g_{lm},  F_{lm})~. \end{equation} $R_{\gamma}$ drops out and the
field equations are: \begin{eqnarray} R_{ab}(g)- \frac{1}{4}R_g~g_{ab} + \kappa~
   T_{ab}(g_{lm}, u^A) =  0~,\nonumber \\ R_g =\frac{c_1}{f'(c_1)}~,~T_{\gamma}= T_g =0~. \end{eqnarray}

In case ii), from (\ref{tracezero}) $R_{\gamma}=c_1=
const$ and we may proceed only if one real solution of $ f'(c_1)~c_1 - 2
f(c_1) = 0~$ does exist and if $f(c_1),~ f'(c_1)$ remain finite. The field 
equations then are \begin{eqnarray} f'(c_1)R_{ab}(g)-\frac{c_1}{4}~g_{ab} + 2\kappa
  T_{ab}(\frac{1}{f'(c_1)}~g_{lm}, u^A) =  0~,\nonumber \\ T_{\gamma}= T_g
  =0~. \end{eqnarray}

In Einstein's theory, $R=0$ follows if the trace of the matter tensor is
vanishing. Here, the larger set of solutions $R = const$ is
obtained.\\

Above, it has been assumed that the matter tensor does not contain covariant 
derivatives; this covers most cases of physical interest. Otherwise,
formidable complications result even when the Einstein-Hilbert
Lagrangian is taken. E.g., if the additional term in the matter Lagrangian is 
$\sim \sqrt{-\gamma} \gamma^{il} \gamma^{km}u_{i;k}~u_{l;m}$
(\ref{det met}) must be replaced by 
$(\sqrt{-\gamma}~\gamma^{ij})_{;k} = f^{ij}_{k}(\gamma^{lm},
u^A, \partial u^A)$ with a particular functional $ f^{ij}_{k}$. Hence, the
elimination of the Lagrangian multiplier will requirec quite an effort.

\section{Comparison with the Palatini method}
\label{section:compalat}
For the Palatini method of variation with variables $g_{ij}$ and
$\Gamma_{ij}^{k}$, the field equations of the f(R)-theory are: \begin{eqnarray}  f'(R) R_{ik}(\Gamma)-\frac{1}{2}
  f(R)g_{ik} = - 2\kappa~T_{ik},~ \label{Palat1}\\ (\sqrt{-g}f'(R)g^{il})_{\parallel l} =0~,\label{Palat2}\end{eqnarray}
where the covariant derivative is formed with the connection
$\Gamma$ and $R=g^{ik}R_{ik}(\Gamma)$. From (\ref{Palat2}) we obtain a
metric $\bar{g}_{ij}$ compatible with the connection $\Gamma$:
\begin{equation}\bar{g}_{ij}=f'(R)~ g_{ij}~,~
  \bar{g}^{ij}=(f')^{-1}(R)~ g^{ij}~, \label{Palatmet} \end{equation}
and the relation between $\Gamma$ and the Levi Civita connection is: 
\begin{equation} \Gamma_{ij}^k \equiv \{ _{ij}^{k}\}_{\bar{g}}=\{
  _{ij}^{k}\}_{g} + \frac{1}{2}\frac{d}{dR}(ln
  f'(R))~[2\delta_{(i}^kR_{,j)}-g_{ij}g^{kl}R_{,l}]~.\label{Palatcon}
  \end{equation} A comparison of (\ref{metric2}) and (\ref{Palatmet}) shows the
difference between $\gamma^{ij}$ and $\bar{g}^{ij}$. With the help of
(\ref{Palatcon}) and (\ref{Palatmet}) we can rewrite 
the tracefree part of (\ref{Palat1}) in terms of the conformally related
metric $\bar{g}_{ij}$ \begin{eqnarray}
  \bar{R}_{ab}(\bar{g})-\frac{1}{4}~ \bar{R}(\bar{g})~\bar{g}_{ab}= 
R_{ab}(g)-\frac{1}{4}~R(g)~g_{ab}-
\frac{f''}{f'}[R_{,i;j}-\frac{1}{4}~g_{ij}  \square R]- \nonumber\\ -
(\frac{f''}{f'}-\frac{3}{2} (\frac{f''}{f'})^2)~
[R_{,i}R_{,j}-\frac{1}{4}~g_{ij}R_{,l}R_{,m}g^{lm}]~. \end{eqnarray}
When bringing the field equations into the form of Einstein's equations, the
result is:\begin{eqnarray}R_{ab}(g)-\frac{1}{2}~R(g)~g_{ab}=- 2\kappa 
T_{ab}({g}) - \frac{f''}{f'}[R_{,i;j}- g_{ij} \square R]- \nonumber\\  -
[\frac{f'''}{f'}-\frac{3}{2} (\frac{f''}{f'})^2]~
R_{,i}R_{,j} + (\frac{f'''}{f'}-\frac{3}{4}
(\frac{f''}{f'})^2)~g_{ij}R_{,l}R_{,m}g^{lm} ~.\label{gradient2}
\end{eqnarray} Again, the trace equation of (\ref{Palat1}), i.e.,
\begin{equation} f'(R) R - 2 f(R) = - 2\kappa~T_g ~, \end{equation} is used 
to eliminate the curvature scalar in favour of the trace of
the matter tensor. This means that the non-minimal coupling to the curvature
scalar and its derivatives will be replaced by a coupling to the {\em
  gradients} of the matter variables contained in $g^{ik}T_{ik}$. 

The remark at the end of section \ref{section:ftheory} for the case of covariant
derivatives in the matter tensor applies here as well.\\

\section{An example: Exponential gravity}
\subsection{New variational principle}
As an example, we now take a recent model for $f(R)$-gravity \cite{Linder09}
with:

\begin{equation}f(R)= -c r (1- e^{-\frac{R}{r}})~, f'=-c
  e^{-\frac{R}{r}}~,\label{expgrav} \end{equation} where $r$ of dimension
$(length)^2$ and $c$, dimensionless, are constants. From (\ref{g-gamma}) $R_g
=-\frac{1}{c}R_{\gamma}e^{\frac{R_{\gamma}}{r}}$. Thus, the inverse
$R_{\gamma}= r^{-1}(R_g)$ can be obtained only numerically. A series expansion 
for $ \frac{R_{\gamma}}{r} << 1$ leads to:\begin{equation} R_{\gamma}=-c R_g -\frac{c^2}{r}R_g^2
  -\frac{3}{2}\frac{c^3}{r^2}R_g^3~\pm \cdot
  \cdot~, \label{expg-gamma} \end{equation}  and \begin{equation}R_g=
  \frac{2\kappa T_g~}{c^2}~[1 - \frac{2\kappa T_g}{rc}+ \frac{2}{3}~
  (\frac{2\kappa T_g}{rc})^2 \pm \cdot  \cdot ] \end{equation}
If the further calculations are restricted {\em to the lowest order} in the 
expansion (\ref{expg-gamma}), with the Einstein tensor $G_{ab}= R_{ab}
-\frac{1}{2}R_g~g_{ab}$  the field equations (\ref{newEineq}) become: 
\begin{equation} 
  G_{ab}(\{_{ij}^{k}\}_g)=
   -\frac{2\kappa}{c^2}~[1- 4\frac{\kappa T_g}{rc}]~T_{ab}~- \frac{\kappa
     T_g}{c^2} ~\frac{\kappa T_g}{rc}~g_{ab}~.\end{equation}
  For a perfect fluid with pressure $p=0$, from (\ref{perfect}) to lowest
  order the equations replacing Einstein's are: 
\begin{equation} R_{ab}(\{_{ij}^{k}\}_g)-\frac{1}{2}R_g~g_{ab}=
  -\frac{2\kappa}{c^2}(1- 4\frac{\kappa \mu}{rc})~ \mu~ u_a u_b -
   \frac{\kappa^2 \mu^2}{rc^3}~g_{ab} \label{newfieldperf}~. \end{equation}
   The result is a variable coupling ``constant'' in the effective matter
   tensor and a variable cosmological term both depending on the energy
   density of matter.

For a homogeneous and isotropic cosmological model with scale factor $a(t)$ and
   flat space sections, (\ref{newfieldperf}) leads to altered Friemann equations:
\begin{eqnarray} (\frac{\dot{a}}{a})^2 = \frac{2\kappa
   \mu}{3c}~(1 -\frac{7}{2} \frac{\kappa \mu}{cr})~,\\ 2\frac{\ddot{a}}{a} + 
(\frac{\dot{a}}{a})^2 = \frac{\kappa^2 \mu^2}{rc^3}~. \label{newFried}\end{eqnarray} Keep in
  mind that $r,c$ are free constants of the model; the velocity of
  light has been put equal to 1 in (\ref{newFried}). As numerical calculations 
would have to be done, and the main aim of this paper is the introduction of a 
new variational principle, we will not comment on this particular model 
(exponential gravity) and the physics following from it.
\subsection{Palatini method}
For exponential gravity as given by (\ref{expgrav}) and for pressureless fluid 
matter, the field equations according to the first equation of (\ref{Palat1}) 
turn out to be \begin{equation} R_{ik}(\Gamma) = \frac{2\kappa}{c}~T_{ik}
  (g,u^A) + g_{ik}~r(1-e^{-\frac{R}{r}}). \end{equation} This does not look
complicated; however the connection $\Gamma$ first  must be expressed by the
conformally related metric $\bar{g}_{ab}$. To the same order of
approximation, the final field equation then can be written as:\begin{eqnarray}
  R_{ab}(g)-\frac{1}{2}~ R(g)~g_{ab}=- 2\kappa \mu
  u_{a} u_{b} -  \frac{2\kappa}{rc}[\mu_{,i;j}- g_{ij} \square \mu]-
  \nonumber\\  - \frac{2\kappa^2}{r^4c^2}~\mu_{,i}~\mu_{,j} +\frac{\kappa^2}{
    r^4c^2}~g_{ij}~\mu_{,l}~\mu_{,m}~g^{lm} ~.\label{expgravgradient}
\end{eqnarray} The effective matter tensor in (\ref{expgravgradient}) depends 
on 1st and 2nd {\em gradients} of the energy density of matter. This shows
that the physics resulting from the two variational principles may be quite
different. The same  can be said with regard to the Einstein-Hilbert
  metric variation used in \cite{Linder09} and e.g., in \cite{SoHuSa07},
  \cite{HuSa07} and the new variational principle.

\section{Concluding remarks}
For physics, a significant difference between the new variational
method presented here and the Palatini method is that non-minimal
coupling of matter and the curvature scalar $R$ occurs by
multiplication with functions of $R$ or the trace of the matter
tensor. In the Palatini method, non-minimal coupling happens via 
the {\em gradients} of the scalar curvature (trace of the matter tensor). A 
conceptual advantage of the new method is that it works within
(pseudo)-Riemannian geometry; metric-affine geometry never does
appear.\footnote{For metric-affine geometry, independent variation of metric
  and connection is mandatory anyway.} When dealing with $R+f(R)$-Lagrangians,
in both approaches a new dimensionful constant is needed whose physical
meaning must be defined. Application to $f(R,~ R_{ab}R^{ab})$ is
unproblematic; here, two new parameters will occur. In general, via
the field equations both curvature invariants can be expressed as
functionals of invariants of the matter tensor. The Einstein-Hilbert Lagrangian
seems to be very robust: now there are at least {\em three} different
methods for a derivation of the Einstein field equations. As the example
treated shows, for more general Lagrangians the variation will lead to
different physical theories. Whether the new variational principle
introduced here, if applied to cosmological models, produces convincing
physics will have to be shown by further studies.

\end{document}